\documentclass[
twocolumn,%
showpacs,preprintnumbers,amsmath,amssymb,
superscriptaddress,unsortedaddress]{revtex4}

\usepackage{graphicx}
\usepackage{dcolumn}
\usepackage{bm}

\newcommand{\om}{\omega}

\newcommand{\eps}{\varepsilon}

\DeclareGraphicsExtensions{.eps,.gif}

\begin{document}


\title{Multiuser quantum communication networks}

\author{Antoni W\'ojcik}
\affiliation{Faculty of Physics, Adam Mickiewicz University,
Umultowska 85, 61-614 Pozna\'{n}, Poland. }
\author{Tomasz {\L}uczak}
\altaffiliation{Corresponding author. Phone: +48 (61) 829-5394, fax:
+48 (61) 829-5315. E-mail: tomasz@amu.edu.pl} \affiliation{Faculty
of Mathematics and Computer Science, Adam Mickiewicz University,
Umultowska 87, 61-614 Pozna\'{n}, Poland. }
\author{Pawe{\l} Kurzy\'nski}
\affiliation{Faculty of Physics, Adam Mickiewicz University,
Umultowska 85, 61-614 Pozna\'{n}, Poland. }
\author{Andrzej Grudka}%
\affiliation{Faculty of Physics, Adam Mickiewicz University,
Umultowska 85, 61-614 Pozna\'{n}, Poland. }
\author{Tomasz Gdala}
\affiliation{Faculty of Mathematics and Computer Science, Adam
Mickiewicz University, Umultowska 87, 61-614 Pozna\'{n}, Poland. }
\author{Ma{\l}gorzata Bednarska}
\affiliation{Faculty of Mathematics and Computer Science, Adam
Mickiewicz University, Umultowska 87, 61-614 Pozna\'{n}, Poland. }

\date{August 8, 2006}%

\begin{abstract}
We study a quantum state transfer between spins interacting with an
arbitrary network of spins coupled by uniform XX interactions. It is
shown that in such a system under fairly general conditions, we can expect a nearly perfect transfer
of states. Then we analyze a generalization of this 
model to the case
of many network users, where the sender can choose which party he
wants to communicate with by appropriately tuning his local magnetic field. 
We also remark that a similar idea can be used  to 
create an entanglement between several spins coupled to the network.
\end{abstract}

\maketitle

\newcommand{\ga}{\gamma}
\newcommand{\la}{\lambda}
\newcommand{\Heff}{H_{\rm eff}}
\newcommand{\ran}{\rangle}
\newcommand{\lan}{\langle}
\newcommand{\hl}{\bar\lambda}


\section{Introduction}

The fact that certain spin systems can be used for transferring quantum
states from a `source' to a `destination' spin  have 
been recently observed and studied by a number of authors 
(see \cite{Bo}--\cite{CCWD}). It has been shown that a
perfect transfer of an arbitrary qubit is possible in spin chains
\cite{CDEL}--\cite{Kay} and some other cases \cite{OL}--\cite{Pater}.
As observed by Bose et al.~\cite{Bo}, 
such a system  can also be
used for creating an entanglement between the source and destination spins.
 In this note we develop  ideas
presented in  \cite{Woj} and prove that one can get a high fidelity 
transfer between two spins weakly coupled to an arbitrary connected network
provided only that they are placed  in a  local
magnetic field we can control. 
We also propose a generalization of this scheme 
and show that  a spin network can be used for communication
with many destination spins, if a source spin tunes his local magnetic 
field to the 'frequency' of an appropriate receiver.

The structure of the note goes as follows. First we describe a
model of Weakly Coupled Spins (WCS) and compute the effective
Hamiltonian of such systems. Then we use our approach to study 
communication between spins coupled to  a spin chain, a cycle,
and, finally, to an arbitrary spin network. In the next section of the
paper we introduce a multi-WCS communication protocol. 
In this model a source spin selects an appropriate `communication
frequency' using which it can effectively communicate with another 
user of the network.  
Finally, we use this scheme to create 
entangled states between network users. 
In particular, we discuss how to get 
an almost perfect Bell state for two spins
and generalized $W$ states in the case of several network users.

\section{WCS Model}

Let us consider two  spins  coupled to a network $G$ of $N$ spins. 
We write the Hamiltonian of a network $G$ as
$$H_G=\sum_{(i,j)} \left( \sigma^{x}_{i}\sigma^{x}_{j} +
\sigma^{y}_{i}\sigma^{y}_{j}\right),$$ 
where we sum over all network edges $(i,j)$. 
A `source' spin $s$ is coupled to a spin
$n_s$ from $G$, and a `destination' spin $d$ is coupled to a $n_d$ from $G$.
The strength of these couplings we denote by $\eps\xi_s$, $\eps\xi_d$,
respectively, where we assume that   
$\xi_s,\xi_d$ are constants independent of $N$ but $\eps$ decreases with $N$. 
Moreover, spins $s$ and $d$ are placed in local
magnetic fields $\omega_{s}$ and $\omega_{d}$, respectively. Then, the
Hamiltonian of the whole system can be written as  $H=H_G + H_{sd}$, where
$H_G$ given above while
\begin{align*}
H_{sd}=\eps\xi_s(\sigma^{x}_{s}\sigma^{x}_{n_s} +
\sigma^{y}_{s}\sigma^{y}_{n_s})&+
\eps\xi_d(\sigma^{x}_{d}\sigma^{x}_{n_d} +
\sigma^{y}_{d}\sigma^{y}_{n_d})\\
&\quad+ \omega_s \sigma_s^z+ \omega_d \sigma_d^z.
\end{align*}
Note that $H$ conserves the total number of excitations.
We shall be mostly interested in the case when there is only one
excitation in the system. Then,  the system remains in the Hilbert space
spanned by vectors $ |n\rangle$, where $n$  denotes the position of the excitation
and takes values  either  $s,d$, or  $1,2,\dots, N$. 
Observe that the Hamiltonian written in this basis is similar to
the network adjacency matrix.

Let $\{\lambda\}$ and $\{|\lambda\rangle\}$ be the sets of the
eigenvalues and the eigenvectors of $H_G$, respectively. Then 
$H=H_0+V$, where 
\begin{align*}
H_{0}&=\omega_s |s\rangle \langle s | + \omega_d |d\rangle \langle d|
+ \sum_{\lambda} \lambda |\lambda \rangle \langle \lambda |,\\
V&=\eps\sum_{\lambda} \left(\xi_s g_{s\lambda} |s\rangle \langle \lambda |
+ \xi_d g_{d\lambda} |d\rangle \langle \lambda | + h.c. \right),
\end{align*}
and $$g_{s\lambda}=\langle s|H_{sd}|\lambda\rangle/(\eps \xi_s),\quad
g_{d\lambda}=\langle d|H_{sd}|\lambda\rangle/(\eps \xi_d)\,.$$

Now, let us consider two cases. 
In the  `resonant' case we have
$\omega_{s}=\omega_{d}=\lambda'$, where $\lambda'$ is one of the
non-degenerated eigenvalues of $G$. Then all terms of the Hamiltonian $H$
corresponding to $\lambda\neq\lambda'$ are of the lower order, and
the evolution of the system takes place  essentially in the 
space $L$ spanned by vectors $|s\rangle$,  $| d\rangle$,  $| \lambda'\rangle$.  
The projection of the Hamiltonian onto $L$
in the basis $\{|s\rangle,   |d\rangle, |\lambda'\rangle\}$ 
can be written as 
\begin{equation}
H^{s \lambda' \!d}_{\rm {eff}} \approx \left(
\begin{array}{ccc}
\lambda' &\eps \xi_sg_{s\lambda'}&0\\
\\
\eps \xi_s^{\ast}g^{\ast}_{s\lambda'}&\lambda' &\eps \xi _d g_{d\lambda'}\\
\\
0&\eps \xi_d^{\ast} g^{\ast}_{d\lambda'} &\lambda'
\end{array}
\right). \nonumber
\end{equation}

In order to obtain an effective Hamiltonian for the  `non-resonant' case, 
where $\omega_s\neq \lambda$ and $\omega_s\neq \lambda$ for all $\lambda$,
i.e., when the local magnetic fields of the source and the
destination spins are not `tuned' to any `natural frequency' of the
network, we follow the approach presented in  \cite{RF}. 
Define a hermitian operator $S$ as
$$S=\eps\sum_{\lambda} \left( \xi_s s_{s\lambda}
|s\rangle \langle \lambda | + \xi_d s_{d\lambda} |d\rangle \langle
\lambda | + h.c. \right),$$ and set $H'=e^{iS}He^{-iS}$.
Note that 
$$H' e^{iS}|\lambda\rangle=   e^{iS}He^{-iS}e^{iS}|\lambda\rangle
= \lambda e^{iS}|\lambda\rangle.$$
Then, expanding $e^{iS}$ and $e^{-iS}$ into power series, we get
$$H'=H_0+V+i[S,H_0]+i[S,V]+\frac{i^2}{2!}[S,[S,H_0]]+\dots.$$
Let us choose $s_{s\lambda}$ and $s_{d\lambda}$ so that the
terms of the first order in $\eps$ vanish, i.e.,
$$V+i[S,H_0]=0.$$
For such a condition  the Hamiltonian $H'$ becomes 
$$H''=H_0+i[S,V]+\frac{i^2}{2!}[S,[S,H_0]]+O(\eps^3).$$
Thus, the projection of $H''$ onto the space generated by 
$\{|s\rangle,  |d\rangle\}$  can be written 
\begin{align*}
H^{sd}_{\rm {eff}} \approx& \left(\omega_s - \eps^2|\xi_{s}|^{2}
\sum_{\lambda}
\frac{|g_{s\lambda}|^2}{\lambda-\omega_{s}}\right)|s\rangle\langle s|  
\\&- \eps^2 \xi_s \xi_d^{\ast} \left(\sum_{\lambda}
\frac{g_{s\lambda}g_{d\lambda}^{\ast}}{\lambda-\omega_{s}} +
\sum_{\lambda}
\frac{g_{s\lambda}g_{d\lambda}^{\ast}}{\lambda-\omega_{d}}
\right)|s\rangle\langle d|  \\
&- \eps^2\xi_s^{\ast}\xi_d \left(\sum_{\lambda}
\frac{g_{s\lambda}^{\ast}g_{d\lambda}}{\lambda-\omega_{s}} +
\sum_{\lambda}
\frac{g_{s\lambda}^{\ast}g_{d\lambda}}{\lambda-\omega_{d}}
\right)|d\rangle\langle s| \\
&+\left(\omega_d - \eps^2|\xi_d|^2 \sum_{\lambda}
\frac{|g_{d\lambda}|^2}{\lambda-\omega_{d}}\right)|d\rangle\langle
d| \,.
\end{align*}

Using the above formulae for $H^{s \lambda' \!d}_{\rm {eff}}$ and  $H^{sd}_{\rm {eff}}$ 
it is not hard to find sufficient conditions under which the state transfer
from $s$ to $d$ occurs. 
In the resonant case, one ought to choose all the off-diagonal case
to be of the same absolute value $\beta\eps$. 
Then, the time of the transfer is, roughly, 
$\frac{\pi}{\sqrt{2}\beta\eps}$.
For the non-resonant case, in order to have a nearly perfect transfer
between $s$ and $d$, one should make  all the diagonal terms of
$H_{\rm{eff}}^{sd}$  equal, and ensure that the off-diagonal terms do not
vanish.
Moreover, if the absolute value of the off-diagonal term
is $\beta'\eps^2$, then the transfer occurs in time 
$T\sim\frac{\pi}{2\beta'\eps^2}$.

Finally, we remark that if we apply the same approach to the case
when both $\omega_d$ and $\omega _s$ are close to a degenerated eigenvalue, we obtain 
an effective Hamiltonian whose projection  $\tilde H_{\rm eff}$ is 
 greater than $3\times3$. Thus,  typically,  no selection of $\xi_s$ and $\xi_d$
 can reduce $\tilde H_{\rm eff}$ to a form which 
 guarantees a perfect transfer.
Thus, in general, we cannot hope to get a perfect transfer by tuning 
to a degenerate eigenvalue  of $H_G$.

\section{Spin networks}

In the paper \cite{Woj} we considered a WCS model 
for spin chains consisting of $N$
spins, where the source and the destination spins $s$, $d$, 
were coupled to the ends of the chain,
and $\omega_s = \omega_d = 0$. We showed that  such a system
admit a high-fidelity transfer of quantum states but
the precise description of this phenomenon depends heavily on the parity of $N$. 
The reason for that is clear when we compute the effective Hamiltonian for such a
system: the case of even $N$ is non-resonant, while for $N$ odd,
when zero is an eigenvalue of the chain, a resonant transmission
occurs.

Now let us consider more general case in which $s$ and $d$ 
are coupled to arbitrary spins of the chain. 
The eigenvectors and the eigenvalues of the chain are given by
$$|\lambda_k\rangle=\sqrt{\frac{2}{N+1}}\sum_{n=1}^{N}\sin\left(\frac{\pi k n}{N+1}\right)
|n\rangle,$$ and
$$\lambda_k=2\cos\left(\frac{\pi k}{N+1}\right),$$
respectively, where $N$ is the chain length and $k=1,2,\dots,N$. 
If $s$ and $d$ are coupled
to the $n_s$th and $n_d$th spin in the chain, respectively, then
$$g_{\alpha\lambda}(n_\alpha)=\sqrt{\frac{2}{N+1}}\sin\left(\frac{\pi k
n_\alpha}{N+1}\right),$$ 
for $\alpha=s,d$.

In the `symmetric case', in which $n_d=N+1-n_s$ and $\xi_s=\xi_d=\xi$, we have 
$g_{s\lambda}(n_s)=g_{d\lambda}(N+1-n_s)$ and, as long as
$\omega_s=\omega_d$, both the resonant and non-resonant cases give
perfect transfer with transfer times of the orders
$\frac{\eps\xi}{\sqrt{N}}$ and $\eps^2\xi^2$, respectively.

\begin{figure}
\scalebox{1.0} {\includegraphics[width=8truecm]{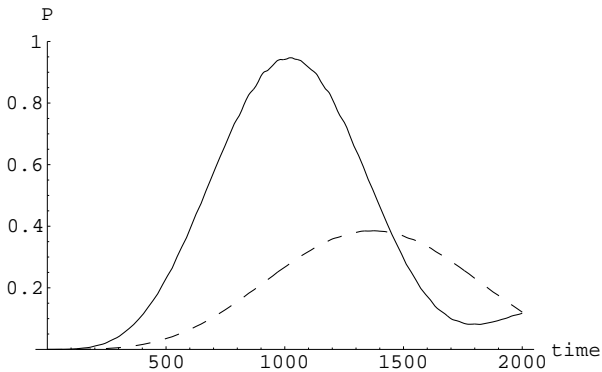}}
\caption{\label{f1} The probability of the excitation of the destination 
spin $d$ as a function of time. 
In this case $N=30$,  $n_s=2$, and $n_d=13$. 
Moreover, we have $\omega_s=\omega_d=\lambda_5$, so the transfer
is resonant.
The dashed
line corresponds to the case $\eps \xi_s = \eps\xi_d=0.01$, the solid one to the
case $\eps\xi_s=0.01$, $\xi_s g_{s\lambda}(2) = \xi_d g_{d\lambda}(13)$.}
\end{figure}

The asymmetric case, when $n_d\neq N+1-n_s$ needs a bit more
attention. If we put $\xi_s=\xi_d$, then $g_{s\lambda}(n_s) \neq
g_{d\lambda}(n_d)$ and the transfer occurs with
probability bounded from above by a constant smaller  than one. 
However, the transfer can be made perfect by switching to
a resonant case. In order to do that one needs to  select the coupling constants
$\xi_s$ and $\xi_d$ in such a way that 
$\xi_s g_{s\lambda}(n_s) = \xi_d g_{d\lambda}(n_d)$, so the condition for a
perfect transfer is satisfied (see Fig.~\ref{f1}).

Another example of communicating through a simple spin network is to
attach two spins to $N$-cycle. The eigenvectors and
eigenvalues for an $N$-cycle are 
$$|\lambda_k\rangle=\sqrt{\frac{1}{N}}\sum_{n=1}^{N}e^{2 \pi i { k
n}/{N}} |n\rangle, $$
and 
$$\lambda_k=2\cos\left(\frac{2 \pi k}{N}\right),$$
respectively.
For a `resonant communication' through $N$-cycle we have only either one or two possible
non-degenerated eigenvectors to choose from, 
namely $|\lambda_N\rangle$ for an odd $N$, 
and $|\lambda_{N/2}\rangle$ and $|\lambda_N\rangle$ for an even $N$.
Moreover, one can check that in this case coupling to a degenerated eigenvector 
cannot lead to a resonant transfer which is nearly perfect. 
This observation is analogous  to a result of Christandl {\it{et al.}}~\cite{CDEL} 
which states that the perfect transfer in the chain with equal
couplings is possible only  for $N=2$ and $N=3$.

Finally, let us consider a general network $G$ with the source spin
$s$ and the destination spin $d$ attached to the nodes $n_s$ and
$n_d$, respectively. A possible transfer of quantum states between
$s$ and $d$ depends on the
eigenvectors $\{|\lambda\rangle\}$ of $G$. Note that each localized
state of the network $|n\rangle$ is a superposition of eigenvectors
$\{|\lambda\rangle\}$. In particular, $s$ and $d$ cannot communicate
unless there exists at least one eigenvector $|\lambda'\rangle$ that
gives non-zero scalar product with both $|n_s\rangle$ and
$|n_d\rangle$; such an eigenvector can be viewed as a 
communication channel. Note however, that for every connected network 
in which all coupling constants are reals, such a vector 
exists. Indeed, if $\lambda'$ is the largest eigenvalue 
of $G$ then, by Perron-Frobenius  theorem, 
$|\lambda'\rangle$ is a vector which corresponds to 
a stationary distribution of a particle in a classical random walk on $G$
and so, for  connected $G$, we have $\langle n| \lambda'\rangle>0$, 
for  $n=1,2,\dots, N$. Observe also that for a connected $G$ the largest eigenvalue 
$\lambda'$ is always non-degenerate, so we can achieve a near perfect transfer 
setting $\om_s=\om_d= \lambda'$, and choosing $\xi_s$ and $\xi_d$ 
so that $\xi_s g_{s\lambda'}=\xi_d g_{d\lambda'}$.
We remark however that for some networks it
is not possible to achieve a nearly perfect transfer in a non-resonant way, 
because for any choice of
$\xi$'s and $\omega$'s either the diagonal terms of the $2
\times 2$ projection of the effective Hamiltonian are not  equal, or 
the off-diagonal terms vanish.

\section{Multiuser quantum network}

\begin{figure}
\scalebox{1.0} {\includegraphics[width=8truecm]{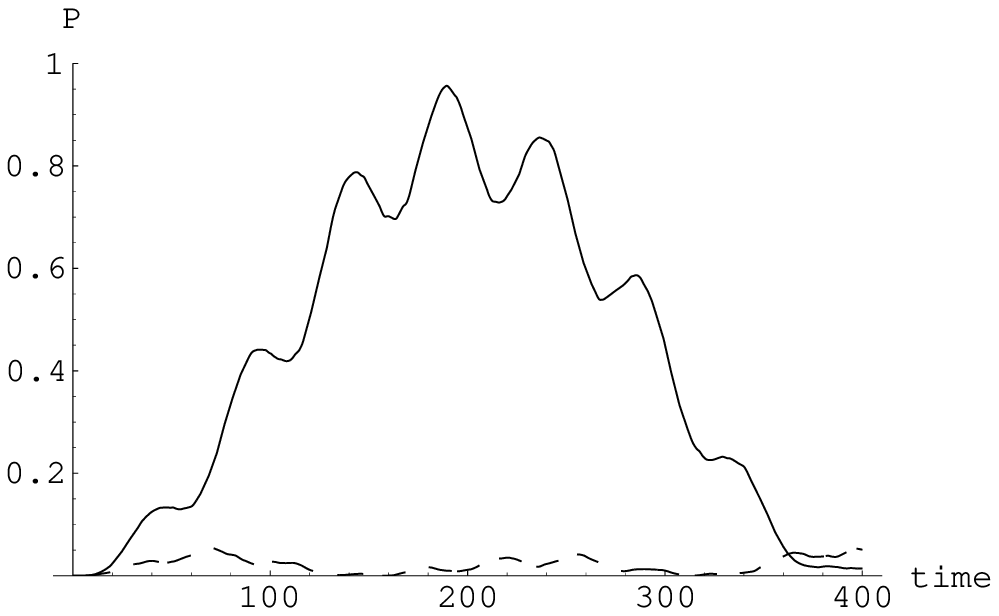}}
{\includegraphics[width=8truecm]{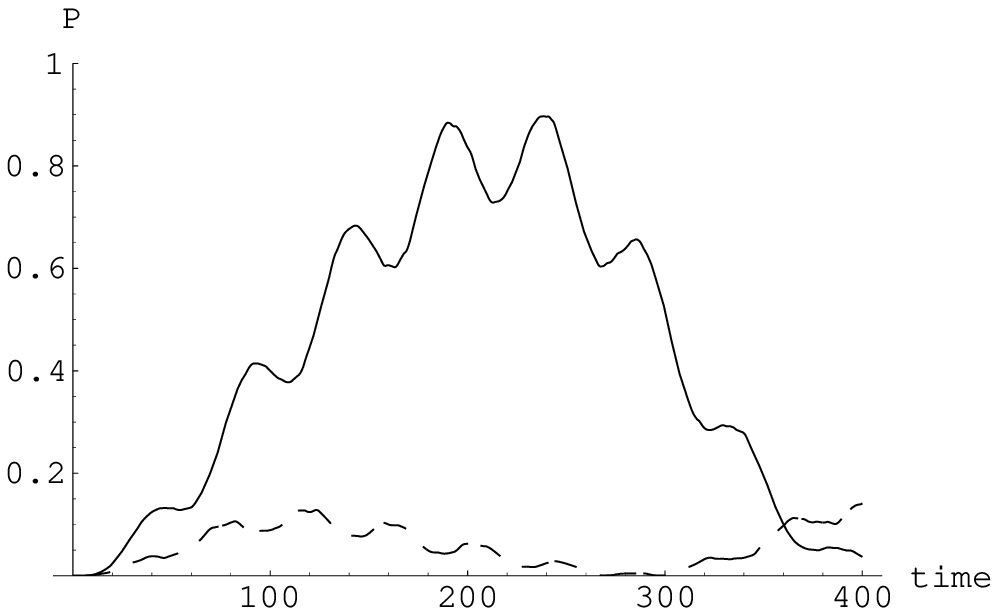}}
{\includegraphics[width=8truecm]{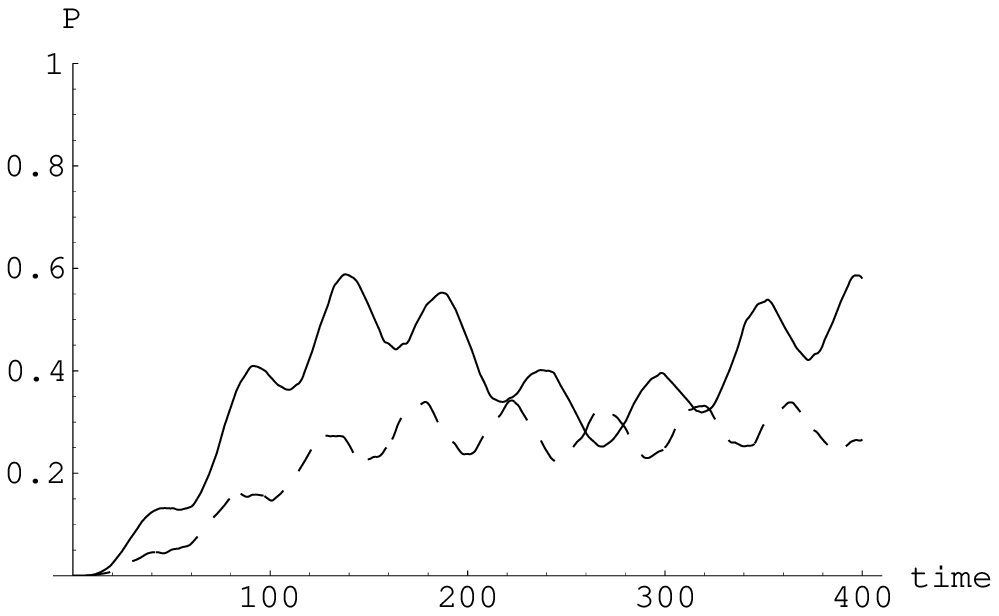}} \caption{\label{f2}
The probability of the state transfer  for an 21-cycle. Three spins
$s$, $u$, and $d$,
are coupled to the 3rd, the 10th, and the 18th  spins of the cycle, respectively.   
The coupling strength is $\eps\xi=0.1$,  and $\omega_s=\omega_d=-0.9$.
The dashed curve represents the population of $u$, the solid one that of $d$.
The top figure corresponds to the case when
$\omega_u=-0.85$, the middle one to $\omega_u=-0.87$ and 
the bottom one to $\omega_u=-0.89$}
\end{figure}

Let us recall that in order to have a nearly perfect transfer
between $s$ and $d$ we should put $\omega_s=\omega_d$; it is also
not hard to check that the fidelity of transfer drops down rapidly
when the difference $|\omega_s-\omega_d|$ grows. This observation
suggests the following multiuser generalization of our communication
protocol. Assume that spins $d_1,d_2,\dots $ are coupled to some
spins of a spin network $G$ and placed in local magnetic fields 
$\omega_{d_1},\omega_{d_2},\dots$, respectively. Then
another spin $s$, coupled to a spin from $G$ as well, can
communicate with any spin $d_k$ by making its own magnetic field
$\omega_s$ equal to $\omega_{d_k}$, and calibrating appropriately
the coupling strength $\xi_s$. To guarantee a high
fidelity of the state transfer from $s$ to $d_k$ we should ensure
that the distance between $\omega_s=\omega_{d_k}$ and the other
frequencies $\omega_{d_i}$, $i\neq k$, as well as between $\omega _s$ and
the eigenvalues $\lambda$ of $G$ (except, perhaps one of them, when
in the resonant case we have $\lambda'=\omega_s$) is large enough.
On the other hand, a large magnetic field $\omega _k$ slows down
the transfer from $s$ to $d_k$. Thus, choosing
frequencies $\omega_1,\omega_2,\dots,$ we should keep in mind both
the fidelity of the transfer  (which decreases with the distance 
between $\omega_i$ and the closest eigenvalue)  and the time of the transmission
(which may increase considerably when $\omega_i$ is large, say, much larger than
 the largest eigenvalue of the network).

We remark that
if in such a protocol another user 
sets his frequency to the communication frequency, the
information will get entangled between him and the intended receiver.
An example of such a disturbance in a system of three spins
coupled to a $21$-cycle is presented on Fig.~\ref{f2}.

 \section{Entanglement generation}                                                      
           
The WCS system can be also used for generation of a perfect entanglement.
Let us consider first a non-resonant communication 
between two users. After the time equal to  a  half of the transfer time,
the state of the system is
$\frac{1}{\sqrt{2}}(|s\rangle+e^{i\phi}|d\rangle)$, where $\phi$ is
some angle which can be easily computed. This state can be also written in the form
$$\frac{1}{\sqrt{2}}\left(|\underbrace{10}_{sd} \overbrace{0 \dots 0}^{network} \rangle
+ e^{i\phi}|\underbrace{01}_{sd} \overbrace{0 \dots 0}^{network}
\rangle \right)=$$
$$=\frac{1}{\sqrt{2}}\left(|01\rangle+e^{i\phi}|10\rangle\right)_{sd}|0\dots
0\rangle_{network},$$ which clearly corresponds to a maximum  entanglement
between $s$ and $d$.

\begin{figure}
\scalebox{1.0} {\includegraphics[width=8truecm]{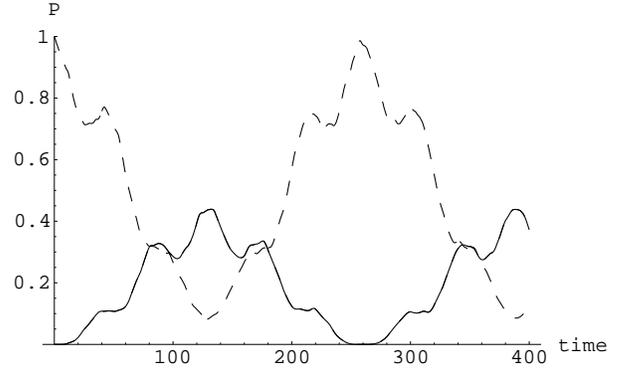}}
\caption{\label{f3} The probability of the excitation of three spins
coupled to a 21-cycle: $s_1$ coupled to the 3rd, $s_2$ to
the 12th, and $s_3$ to the 15th spins of the cycle. 
The dashed line represents the excitation of $s_1$,
the solid one that of $s_2$ and $s_3$.
Here we set $\omega_{s_1}=\omega_{s_2}=\omega_{s_3}=-0.9$;
 the  coupling strength is  $\eps\xi=0.1$. 
The $W$ state is obtained at times when two lines meet.}
\end{figure}

In order  to obtain $W$ state one has to  consider three spins
and a non-resonant effective Hamiltonian of the form
\begin{equation*}
H^{\rm multi}_{\rm eff} \approx \left(
\begin{array}{ccc}
\gamma & \alpha & e^{i\varphi}\alpha\\
\\
\alpha^{\ast} & \gamma & \beta\\
\\
e^{-i\varphi}\alpha^{\ast} & \beta^{\ast} & \gamma
\end{array}
\right). 
\end{equation*}
Let us set the initial conditions as
$|\Psi(0)\rangle=|s_1\rangle$. 
After an easily computable time,  the state of the system 
becomes a  $W$ state
$\frac{1}{\sqrt{3}}\left(|s_1\rangle+e^{i\phi}|s_2\rangle+e^{i\theta}|s_3\rangle\right)$,
for some $\phi$ and $\theta$ (cf.~Fig.~\ref{f3}).
               
We can also use a resonant transfer to get an entanglement for $m\ge 2$ users.
To this end we apply the following procedure.
First, one user $s_1$ in an initial state
$|\Psi(0)\rangle=|s_1\rangle$  couples to a non-degenerated
eigenvalue $\lambda'$ of the network and waits until the 
system evolves into $|\lambda'\rangle$. 
This evolution is described by the following effective Hamiltonian
$$H_{\rm eff}^{1}=\lambda'|\lambda'\rangle\langle\lambda'|+\lambda'|s_1\rangle\langle s_1|+
\left(\eps_{\xi_1}g_{s_1\lambda'}|s_1\rangle\langle\lambda'|+h.c.\right).$$
Then  all $m$ users couple to excited $|\lambda'\rangle$ state via
the effective Hamiltonian
\begin{align*}
H_{\rm eff}^{2} =& \lambda' |\lambda'\rangle\langle \lambda'| +
\sum_{i} \lambda' |s_i \rangle\langle s_i|  \\
&+\left(\eps\sum_{i}^{M} \xi_{i} g_{s_i\lambda'} |s_i\rangle\langle
\lambda'| + h.c. \right), 
\end{align*}
where $\xi_i g_{s_i\lambda'}=\xi_{j} g_{s_j\lambda'}$  for all
pairs $\{i,j\}$, and wait until the system evolves to the state
$$\frac{1}{\sqrt{M}}\sum_{j=1}^M e^{i\varphi_j}|s_j\rangle.$$ 
Note, that this method is similar to the dynamics of the  spin star network
presented in \cite{CCWD}. Observe also that one
can get rid of all relative phases by local one qubit operations.

  \section{Summary} 

In the paper we generalized a number of earlier results on 
quantum information transmission between a source spin and 
a destination spin using a simple spin network.
We  show that a near-perfect state transfer between two spins 
is possible through a large class 
of networks, provided that we can control local magnetic field in which 
the source and the destination spins are embedded. We also point out
that a source spin can choose the destination spin from a number of
`users' of the network by an appropriate `tuning', i.e., by
carefully selecting  its local magnetic field. The very same 
mechanism can be used to generate a multispin  entanglement state.

\section{Acknowledgments}

The authors  wish to thank the State Committee for Scientific Research
(KBN) for its support: 
 A.W.\ and A.G.\ were supported by grant 0~T00A~003~23 
T.\L., T.G.\ and M.B.\ by grant 1 P03A 025 27.

\end{document}